\documentclass[10pt,a4paper,twocolumn]{article}
\usepackage{fukugo}
\usepackage{amsmath}
\usepackage{amsfonts}
\usepackage{amssymb}
\begin{document}
%
\title{Masses and mixing of neutrinos in grand unified SO(10) model}
%
%
%

\author{{\large{Noriyuki OSHIMO}} \medskip \\ 
  \it{ Department of Physics and Chemistry,} \\ 
  \it{ Institute of Humanities and Sciences,       } \\ 
  \it{ Ochanomizu University                             }
       }
%
%
%
\maketitle
\pagestyle{empty}
\thispagestyle{empty}
\setlength{\baselineskip}{13pt}
%
%

%
%
\section{Introduction}

     Non-vanishing but extremely small masses for neutrinos 
are deduced from the measurements of the solar neutrino deficit  
and the atmospheric neutrino anomaly.  
If the neutrino masses does really not vanish, 
the existence of physics beyond the standard model is 
immediately implied.  
Then, grand unified theories (GUTs) are most plausible.   
However, it has also been observed that 
the Maki-Nakagawa-Sakata (MNS) matrix, which describes  
the generation mixing among the leptons similarly to the 
Cabibbo-Kobayashi-Maskawa (CKM) matrix for the quarks, 
has unsuppressed off-diagonal elements.  
The off-diagonal elements of the CKM matrix are small.  
This difference between the quarks and the leptons poses 
a severe constraint on GUT models, since the Higgs couplings 
of the quarks and the leptons are closely related to each other.    

\smallskip

     We propose a GUT model based on SO(10) and supersymmetry [NO1].  
The masses and mixings of the quarks and the leptons 
are well described simply by introducing a Higgs superfield 
belonging to $\bf 120$ representation, in addition to ordinarily 
contained $\bf 10$ and $\overline {\bf 126}$ representations.
Below the GUT energy scale, the model is the same as the minimal
supersymmetric standard model except the inclusion of 
dimension-5 operators.  
These new terms lead to small Majorana masses for the left-handed neutrinos
after electroweak symmetry is broken down.

%
\section{Model}

     The grand unified group of our model is SO(10).
Its spinor $\bf 16$ representation contains all
the quark and lepton superfields of one generation,
both left-handed and right-handed components.  
The right-handed neutrinos are thus naturally incorporated.
For the Higgs superfields which give masses to the quarks and leptons, 
one superfield is introduced for each of $\bf 10$, $\bf 120$, 
and $\overline{\bf 126}$ representations.
These three representations are all that can couple to the direct 
product of ${\bf 16}$ and ${\bf 16}$.  

\medskip

     The representation $\bf 10$ contains SU(2) doublets $H^{\bar 5}_{10}$ 
and $H^5_{10}$ with hypercharges $Y=-1/2$ and $Y=1/2$, respectively, 
where upper indices denote transformation properties under SU(5).  
In the representation $\bf 120$ there are four SU(2) doublets: 
$H^{\bar 5}_{120}$ and $H^{\overline{45}}_{120}$ with $Y=-1/2$ and 
$H^5_{120}$ and $H^{45}_{120}$ with $Y=1/2$.  
The representation $\overline{\bf 126}$ involves SU(2) doublets   
$H^{\overline{45}}_{\overline{126}}$ with $Y=-1/2$ and 
$H^5_{\overline{126}}$ with $Y=1/2$, an SU(2) triplet 
$H^{15}_{\overline{126}}$, and an SU(2)$\times$U(1) singlet 
$H^1_{\overline{126}}$.  

\medskip

     The Higgs doublet superfields $H_1$ with $Y=-1/2$ and $H_2$ 
with $Y=1/2$ responsible for electroweak symmetry breaking are given 
by linear combinations of possible SU(2)-doublets with the same 
hypercharges, which we can write as   
$$
\begin{aligned}
H_1 &= (C_1^\dagger)_{11}H^{\bar 5}_{10}
          +(C_1^\dagger)_{12}H^{\bar 5}_{120}
          +(C_1^\dagger)_{13}H^{\overline{45}}_{120} \\
  &+(C_1^\dagger)_{14}H^{\overline{45}}_{\overline{126}}+...,  \\
H_2 &= (C_2^\dagger)_{11}H^5_{10}
          +(C_2^\dagger)_{12}H^5_{120}
          +(C_2^\dagger)_{13}H^{45}_{120}    \\
  &+(C_2^\dagger)_{14}H^5_{\overline{126}}+...,
\end{aligned}
$$
where $C_1$ and $C_2$ represent unitary matrices.  
Some components of $H_1$ and $H_2$ may belong to the representations
different from $\bf 10$, $\bf 120$, and $\overline{\bf 126}$,
which are expressed by the ellipses.  
The other linear combinations of SU(2) doublets should be sufficiently 
heavy below the GUT energy scale to satisfy the unification of the 
gauge coupling constants for SU(3)$\times$SU(2)$\times$U(1).  
The scalar component of $H^1_{\overline{126}}$ is assumed to have 
a vacuum expectation value (VEV) $v_S/\sqrt{2}$ of the order of 
a GUT energy scale, giving large masses to  
right-handed neutrinos and sneutrinos.  
We take $H^{15}_{\overline{126}}$ for heavy enough not to develop 
a non-vanishing VEV.

\medskip

     The superpotential relevant to the quark and lepton masses
are given, in the framework of SU(3)$\times$SU(2)$\times$U(1), by
$$
\begin{aligned}
W &= \eta_d^{ij} H_1 Q^iD^{cj} + \eta_u^{ij}H_2 Q^iU^{cj}
 + \eta_e^{ij} H_1 L^iE^{cj}  \\
 &+ \frac{1}{2}\kappa^{ij}H_2 L^iH_2 L^j + {\rm H.c.},
\end{aligned}
$$
where $Q^i$, $U^{ci}$, and $D^{ci}$ denote the quark superfields 
and $L^i$ and $E^{ci}$ represent the lepton superfields  
in self-explanatory notations,  
with $i$ being the generation index.
The dimension-5 terms proportional to $\kappa$ are induced by exchanging  
the heavy right-handed neutrinos and sneutrinos.  

\medskip

     The coefficients $\eta_d$ and $\eta_u$ become symmetric concerning 
generation indices,   
if the $(\sqrt{3}H^{\bar 5}_{120}+H^{\overline{45}}_{120})/2$ component 
in $H_1$ and the $H^{45}_{120}$ component in $H_2$ can be neglected.  
Taking a generation basis in which the coefficient matrix for the up-type
quarks is diagonal, they are written as  
$$
\begin{aligned}
\eta_u &= \eta^D_u,   \\
\eta_d &=V_{CKM}^*\eta^D_dV_{CKM}^\dagger,
\end{aligned}
$$
where $\eta^D_u$ and $\eta^D_d$ represent diagonal matrices, and 
$V_{CKM}$ stands for the CKM matrix.  
The coefficients for the leptons are given by 
$$
\begin{aligned}
\eta_e &= -\frac{3r_1+r_4}{r_1-r_4}\eta_d +\frac{4}{r_1-r_4}\eta_u
            +4\tilde\epsilon,  \\
\kappa &= -\eta_\nu\left(M_{\nu_R}\right)^{-1}\eta_\nu^T, \\
\eta_\nu &= -\frac{4r_1r_4}{r_1-r_4}\eta_d +\frac{r_1+3r_4}{r_1-r_4}\eta_u
             + 2r_2\tilde\epsilon,  \\
M_{\nu_R} &= \frac{2\sqrt{3}v_S}{(C_1)_{41}}
     \left[\frac{r_1}{r_1-r_4}\eta_d -\frac{1}{r_1-r_4}\eta_u\right], \\
 & \tilde\epsilon = (C_1)_{21}\epsilon,   \\ 
&r_1 = \frac{(C_2)_{11}}{(C_1)_{11}}, 
    \quad  r_2=\frac{(C_2)_{21}}{(C_1)_{21}},
\quad r_4=\frac{(C_2)_{41}}{(C_1)_{41}},
\end{aligned}
$$
where $\epsilon$ denotes the coefficients for the couplings 
${\bf 120}\times{\bf 16}\times{\bf 16}$.  
Owing to the group structure, $\epsilon$ is antisymmetric.  
The parameters for the leptons are expressed by those for 
the quarks and the additional six parameters $r_1$, $r_2$, $r_4$, and 
$\tilde\epsilon$.  

%
\section{Numerical results}

     The independent parameters at the GUT energy scale are given by
the diagonal matrices $\eta^D_u$ and $\eta^D_d$, the
CKM matrix $V_{CKM}$, the ratio $r_1$, $r_2$, and $r_4$,
the antisymmetric matrix $\tilde\epsilon$, and the right-handed
neutrino mass scale $v_S/(C_1)_{41}$.
At the electroweak energy scale, the eigenvalues of
$\eta_u$, $\eta_d$, and $\eta_e$ are known experimentally,
if the ratio $\tan\beta$ of the VEVs for $H_1$ and $H_2$ is given.
The CKM matrix has been measured.
The quantities obtained experimentally for the neutrinos are
the mass-squared differences and the MNS matrix.

\medskip

     The matrices $\eta_u$, $\eta_d$, $\eta_e$, and $\kappa$ 
evolve depending on the energy scale.  
To discuss phenomena at the electroweak energy scale, 
we make analyses with the aid of the renormalization group equations for 
$\eta^D_u$, $\eta^D_d$, $\eta^D_e$, $\kappa^D$, $V_{CKM}$, and $V_{MNS}$.   
The observed quantities at the electroweak energy scale have to be 
accommodated by suitable values of the parameters at the GUT energy scale.  

\medskip

     The numerical analyses show that the measurements of the
neutrino masses and the MNS matrix are explained well
in certain regions of the parameter space [NO2].
The experimental results for the quark and charged lepton masses and those 
for the CKM matrix, though they stringently constrain the model parameters, 
can also be satisfied.  
The Higgs bosons in $\bf 120$ make the mixing structures different
between the quarks and the leptons.  
The $\overline {\bf 126}$ representation is the origin of the 
extreme smallness of the neutrino masses.  
Without invoking contrived schemes, the masses and mixings of the quarks and 
the leptons can be accommodated consistently by a simple extension of 
the minimal SO(10) model.   

\medskip

     This work is supported in part by the Grant-in-Aid for
Scientific Research on Priority Areas (No. 14039204) from the
Ministry of Education, Science and Culture, Japan.

%
%

\end{document}